\documentclass[12pt,a4paper]{article}
\usepackage{amsmath}
\usepackage{graphicx}
\usepackage{times}
\usepackage{soul}
\usepackage{cite}
\begin{document}
\begin{titlepage}

\title{ Energy evolution of the overlap functions: increasing ratio of $\sigma_{el}(s)/\sigma_{tot}(s)$ and black ring emergence}
\author{ S.M. Troshin, N.E. Tyurin\\[1ex]
\small   NRC ``Kurchatov Institute''--IHEP\\
\small   Protvino, 142281, Russia\\}

\normalsize

\date{}
\maketitle

\begin{abstract}
We consider  two  possible options of the energy dependency of the elastic and inelastic overlap functions. Those  correspond to saturation of the black disk limit (BEL effect) and to the unitarity  saturation (REL effect)  at $s\to\infty$. Relation of the REL effect  to  increase of the ratio $\sigma_{el}(s)/\sigma_{tot}(s)$  and emergence of black ring picture at the LHC   is underlined.

\end{abstract}
\end{titlepage}
\setcounter{page}{2}
\section*{Introduction}
The note considers the energy dependency of the overlap functions based on the available experimental data with the aim of obtaining  hints for  the asymptotics of the soft hadron interactions. Relation of   exceeding the black disc limit  with  increase of the ratio $X(s)\equiv \sigma_{el}(s)/\sigma_{tot}(s)$  at the LHC energies  is underlined. The relevant tool  is to be reconstruction of the impact--parameter dependent quantities from the differential cross--section of elastic scattering.  This reconstruction has ambigutes related to the large $-t$ region. Those are discussed thorougly in \cite{l1,l2} where the model-independent approach based on the L\'evy imaging method has been proposed \cite{l1}. This analysis reveals a presence of the black-ring effect at the energy of $\sqrt{s}=13$ TeV. We discuss a qualitative possibility to further minimize ambiguity of such reconstruction focusing on the energy dependency of the probability distributions in the elastic and inelastic head--on collisions, i.e. at  zero impact parameter value. Their energy dependencies constitute a central subject of the note. 

 We list first  several   results   from the CERN ISR energies.
The ISR measurements demonstrated that the models with Geometrical Scaling (GS)\footnote{Geometrical scaling implies reduction of the scattering amplitude dependence on the two independent variables -- energy and impact parameter $b$-- to  dependence on the single variable defined as a ratio of the impact parameter to the interaction radius. Note, that the position operator (impact parameter $b$) is deprived of
	its noncommutativity with the Hamiltonian when the energy increases \cite{webb}.} are in a good agreement with the data \cite{amaldi}. GS implies, in particular, an energy--independence of the elastic to total cross--sections ratio $X(s)$. The value for the ratio $X(s)$ has been estimated to be around $0.18$ across the CERN ISR energy range (covers  about 40 GeV of $\sqrt{s}$variation).

Assumption on the crossing--even dominance in the scalar elastic scattering amplitude combined with the data from $S\bar p p S$ collider
have led to prediction of rising  $X(s)$ from $0.18$ to $0.21$ \cite{henziv,henzi} between the ISR energy and the energy of $\sqrt{s}=546$ GeV . It was claimed that GS is violated and the above increase is a result of the ``blacker--and--edgier'' components of the so called BEL effect. This abbreviation stands for the three terms: Blacker, Edgier, and Larger.  Those describe an energy evolution  of the inelastic interactions region   in the impact parameter space.  Such evolution has  been confirmed by CDF collaboration at Tevatron where increase of $\mbox{Im} f(s,0)$ from the value of
$0.36$ at the CERN ISR to the value of $0.492\pm 0.008$ at $\sqrt{s}= 1800$ GeV has been established  \cite{giro}.  The function $f(s,b)$ is  the Hankel (Fourier-Bessel) transform of the scattering amplitude $F(s,t)$. 

Trend on the rise of the ratio $X(s)$ continues up to the maximal energy of the LHC $\sqrt{s}=13$ TeV where this ratio obtained the value of $0.28$  \cite{rat, drem}.  This ratio can vary in the range from $0$ to $1$ and a comprehensive discussion of possible scenarios for the asymptotic  values of $X(s)$  is given in \cite{fagun}. Evidently, increase of the relative number of elastic events correlates with respective decrease of the relative number of inelastic events.

Unitarity in the impact parameter representation relates the elastic and inelastic  overlap functions introduced by Van Hove \cite{vh} with $h_{tot}(s,b)\equiv \mbox{Im} f(s,b)$ in the following way
\begin{equation}
h_{tot}(s,b)=h_{el}(s,b)+h_{inel}(s,b).
\end{equation}	
The respective elastic, inelastic and total cross--sections are determined by  the integrals	of $h_i$:
\begin{equation}
\sigma_i(s)=8\pi\int_0^\infty bdb h_i(s,b)
\end{equation}	
with $i=tot,el,inel$. Unitarity relation for the overlap functions depending on $s$ and $t$ variables is 
\begin{equation}
	H_{tot}(s,t)=H_{el}(s,t)+H_{inel}(s,t).
	\end{equation}	
The overlap functions functions $H_i(s,t)$ are the Fourier-Bessel (i.t. the particular case of the Hankel) transforms of the overlap functions $h_i(s,b)$ in the impact parameter space:
\begin{equation}
H_i(s,t)=\frac{s}{\pi ^2}\int_0^\infty bdb h_i(s,b)J_0(b\sqrt{-t}).
\end{equation}

The limiting value  the amplitude tends to (we disregard an oscillating dependence) is not clear nowadays. 
We restrict our consideration by the two cases: the most common one corresponding to the maximal inelastic channels contribution to the unitarity with the asymptotics (black disc limit) 
\begin{equation}
X(s)\to 1/2
\end{equation}
and to the reflective scattering mode which implies unitarity saturation at asymptotics with the limiting ratio
\begin{equation}
X(s)\to 1\
\end{equation}
at $s\to\infty$ .
One should note that the current experimental values of $X(s)$ are still  far from the both limits. 
But, situation is drastically different in the impact parameter space. It has been shown in several papers \cite{l2,b1,b2,b3} that there is no room for the ``Blacker'' component of the BEL effect at the LHC energies\footnote{At the LHC energy range, the unitarity relation  at the impact parameters  range $0\leq b\simeq 0.4$ fm  gives \cite{l2,b1,b2,b3}
		$(\mbox{Im}f(s,b)-{1}/{2})^2 +(\mbox{Re} f(s,b))^2\simeq 0$.}.
Inelastic overlap	function  $h_{inel}(s,b)\equiv \sigma_{inel}(s,b)\equiv d\sigma_{inel}/4\pi db^2$ has reached its maximal value allowed by unitarity in the head--on collisions \cite{b1,b2,b3}, and
\begin{equation}\label{neg}
\partial h_{inel}(s,b)/\partial s \leq 0
\end{equation}
at fixed $b=0$,
but the ratio 
$X(s)$ continues to rise with energy at the LHC while   Eq. (\ref{neg}) takes place. 
{\it Thus, the reason of  $X(s)$ growth  is different from being a consequence of the ``blacker--and--edgier'' component of the BEL effect at these energies (cf. \cite{henziv,henzi})}.

We suggest here to treat this further increase of the elastic to total cross--sections  ratio as a result of transition
to the reflective scattering mode at the LHC energy range when the elastic scattering amplitude in the impact parameter space crosses the black disc limit exceeding it. 

We neglect for simplicity the effects of a real part of the amplitude making replacement $f\to if$.   It is just an approximation based on its small magnitude.  However, the real part  account does not affect conclusions of this note \cite{anal}.

The region of small impact parameters (vicinity of $b=0$) is the most sensitive  to appearance of the reflective scattering  mode. Physical interpretation of this mode has been discussed in \cite{physl21} and \cite{jpg}.

\section{Transition to the reflective scattering mode}

This mode implies unitarity saturation at asymptotics and is relevant for the  LHC energies and beyond. The reflective scattering mode appears in the region of  the energies  and impact parameters where the amplitude $f$ has the values in the range of $1/2 < f \leq 1$. 
There are  indications of its appearance at the LHC
\cite{mart} despite 
the   analysis of the data at the energy of $\sqrt{s}=13$ TeV  are still not yet convergent \cite{totem, dremin}.

Elastic and inelastic overlap functions can be expressed through the scattering amplitude in the following forms
\begin{equation}\label{el}
	h_{el}(s,b)= f^2(s,b),
\end{equation}
\begin{equation}\label{inel}
	h_{inel}(s,b)= f(s,b)[1-f(s,b)].
	\end{equation}
The relation between derivatives of the elastic and inelastic overlap functions
\begin{equation}\label{cent}
	\frac {\partial h_{el}}{\partial s}=\left[ \frac{1-S}{S}\right]\frac{\partial h_{inel}}{\partial s}
\end{equation}
  ($S$ denotes the elastic $S$--matrix scattering element,  $S\equiv S(s,b)$) points to a strong increase of elastic overlap function in case of small negative deviation of the inelastic overlap function at small values of the impact parameters 
($S$ is small and negative)\footnote{The derivatives $ {\partial h_{el,inel}}/{\partial s}$  at $b=0$ provide a nontrivial information on energy dependence contrary to the derivatives ${\partial h_{el,inel}}/{\partial b}$ at $b=0$, which are vanishing at any  energy as a consequence of the Hankel transformomation.}. Negative $S$ makes
increase of  $\sigma_{el}(s,b)$ in vicinity of  small  impact parameters faster than the corresponding increase  of $\sigma_{tot}(s,b)$ with energy, since 
\begin{equation}\label{centn}
	\frac {\partial h_{el}}{\partial s}=\left( {1-S} \right)\frac{\partial f}{\partial s}.
\end{equation}
It gives qualitative explanation of the  observed growth of
$X(s)$  in the LHC energy range. 

It should be noted that the following relation has to be valid for the inelastic overlap function $H_{inel}(s,t)$ in the reflective scattering mode
\begin{equation}
I(s)\equiv \int_0^\infty \sqrt{-t} d\sqrt{-t}\left[ s\frac{\partial H_{inel}(s,t)}{\partial s}-H_{inel}(s,t)\right]<0,
\end{equation}
as a consequence of the fact, that the function
\begin{equation}
h_{inel}(s,0)=\frac{\pi ^2}{s}\int_0^\infty H_{inel}(s,t)\sqrt{-t}d\sqrt{-t}
\end{equation}
should decrease with energy in the reflective scattering mode and be an energy--independent positive constant in the case of black disk limit saturation, i.e.
when  the BEL effect stops at $b=0$, the function $I(s)$ should not depend on $s$ and turns to be $0$. A simple exponential approximation for the differential cross--section of the elastic proton scattering has been used in \cite{dre} for estimation of $h_{inel}(s,0)$ for the energies up to $\sqrt{s}=7$ TeV. Extrapolation to higher energies  \cite{white} has been performed on the model base in \cite{kfk}. This leads to extrapolation with $I(s)\simeq 0$ saturating black disk limit. Another option corresponds to $I(s)<0$ at the LHC energies and beyond. It can be illustrated by the $U$--matrix unitarization  scheme \cite{savr} which includes both the absorption and reflection and ensures fulfillment of inequality $f\leq1$ as it is required by unitarity at all the energies. In this scheme, the  elastic scattering amplitude  is represented in the rational form \cite{savr,int}:
\begin{equation}\label{uma}
f(s,b)=u(s,b)/[1+u(s,b)].
\end{equation}
 In the nonrelativistic potential scattering this representation of the amplitude was established in \cite{bg}.

The form of the input amplitude $u$  in the relativistic case  can be invoked from  a phenomenological models (QCD--inspired, Regge model, geometrical model or other).  General principles of the theory provide a framework under the choice of the model forms.

Currently, the knowledge of the QCD Lagrangian provides a little help in the nonperturbative region  where confinement and spontaneous chiral symmetry breaking presumably play an important role. 
We use geometrical model for  construction of the input amplitude  since it is transparent and allows to use considerations on the form of the interaction region and hadron structure, the general arguments in favor of the factorized  form of $u(s,b)$ and its power-like energy dependence are given in \cite{epl,sym}:
\begin{equation}
u(s,b)=g(s)\omega(b),
\end{equation}
where $g(s)$ is positive and $\sim s^\lambda$ with positive exponent $\lambda$ at high energies and $\omega(b)=\exp(-\mu b)$. 
The use of the Regge model with Pomeron intercept $\alpha_P(0)>1$\footnote{The nowadays popular Odderon should have  intercept $\alpha_O(s)\leq \alpha_P(0)$. This inequality is due to requirements of  various unitarization mechanisms \cite{fin,martyn} and  Odderon contribution is not leading and is not mentioned therefore in the  context of the present discussion (see \cite{br} and references therein in this regard). Moreover, arguments in favor of the strict inequality can be found in \cite{fin}. The nonmonotonic energy dependence of the total cross--sections is not foreseen while continuing increase of the total cross-ssection is under consideration.} makes no qualitative difference for the energy dependence of the inelastic overlap function $h_{inel}(s,0)$. For example, hard Pomeron parametrization has been used as an input for unitarization in \cite{cud}.
The inelastic overlap function  at $b=0$ has the form
\begin{equation}
h_{inel}(s,0)=u(s,0)/[1+u(s,0)]^2	
=g(s)/[1+g(s)]^2=sf'(s,0)/\lambda.
\end{equation}
Thus, the  function $h_{inel}(s,0)\sim s^{-\lambda}$ at $s\to\infty$ and
respectfully the function $I(s)$ is positive at $g(s)<1$, it is equal to zero at $g(s)=1$ (maximal absorption) and becomes to be negative with growing energy, when  $g(s)>1$ (note that $\omega(0)=1$). Indeed, 
\begin{equation}
I(s)=[1-g(s)]g'(s)/[1+g(s)]^3
\end{equation}
	and the function $g'(s)$ is positive since $\lambda>0$. The experimental data imply that $g(s)\simeq 1$ at the LHC energy range\cite{b1,b2,b3}. This energy region  is critical for extrapolation of  $I(s)$ behavior to higher energies.
		The inelastic overlap function $h_{inel}(s,0)=h_{inel}$ performs  loop variation  with the increase of energy  $s_i\to s_m\to s_f$; it varies as $h_{inel}^i\to 1/4\to h_{inel}^f$ and $h_{inel}^f=h_{inel}^i$. This loop variation is due to the symmetry $h_{inel}=f(1-f)$\footnote{In fact, the exact symmetry relates to $\mbox{Im}f$ in general case (when real part of the scattering amplitude is taken into account).} under replacement $$f\leftrightarrow1-f$$ resulting from unitarity. Such variation of $h_{inel}$ cannot be realized when the absorptive scattering mode exists only ($f\leq 1/2$) and symmetry under replacement  $f\leftrightarrow1-f$ can be realised at one energy value only, i.e. when $f=1/2$. It indicates that  exclusion of the reflective scattering mode ($f> 1/2$), i.e.  when $h_{el}$ exceeds $h_{inel}$, has little ground.
	Achieving the maximum of $h_{inel}=1/4$ and further  decrease of this function does not depend on the particular form of the $U$--matrix, and is a consequence of the asymptotic unitarity saturation, which corresponds to  implementation of the principle of the maximal strength of strong interactions \cite{frau}.  Maximum of  $h_{inel}(s,0)$ happens at the LHC energy range, i.e. $$s_m=s_{LHC}.$$
 The phenomena of further deepening  at $b=0$ can be associated with the hypothesis of  effective account    of the phase randomization  mechanism of the multiparticle production amplitudes    and increase of the mutual cancellation of these amplitudes'  contributions  which increases with energy. This hypothesis is inspired \cite{pred} by  the explanation of the diffraction peak origin stemming from  a shadow nature of  elastic scattering 
 amplitude\footnote{However, this particular mechanism of the diffraction peak formation is not universal and valid for shadow scattering only. It implies an absence of such peak when the reflective scattering mode dominates.}.

It would be interesting to study experimental energy evolution of the function $I(s)$ to detect  a transition from the BEL ($I(s)>0$) to REL regime ($I(s)<0$).  Here capital ``R'' stands for adjective ``reflective''. The advantage of the function $I(s)$ use is in its integral nature.  It is mostly relevant for the large $-t$ region with small value of the differential cross--section.
Similar function $E(s)$ defined as 
\begin{equation}
	E(s)\equiv \int_0^\infty \sqrt{-t} d\sqrt{-t}\left[ s{\partial H_{el}(s,t)}/{\partial s}-H_{el}(s,t)\right].
\end{equation}
should be vanishing in case of the black disk limit saturation and remain positive in case of  transition to the reflective scattering at the LHC, (cf. Eq. (\ref{centn}).
The relation between the functions $E(s)$ and $I(s)$ follows from Eq. (\ref{cent} at $b=0$):
\begin{equation}\label{cent1}
	E(s)=[{(1-S)}/{S}]I(s).
\end{equation}
The both functions $I$ and $S$ are negative in the reflective scattering mode ($S<0$) and the function $E$ is nonnegative in both absorptive ($S>0$) and reflective modes.
\section{Saturation of unitarity and black disk limits}
Saturation of unitarity limit (REL) corresponds to the following behavior of the scattering amplitude in the impact parameter space  at $b=0$:
\begin{equation}
f(s,0)\to 1
\end{equation}
at $s\to\infty$.
Saturation of the black disk limit (BEL) implies that asymptotically
\begin{equation}
	f(s,0)\to 1/2
\end{equation}
from below.
Thus, the following dependencies should take place at $s\to\infty$:
\begin{equation}\label{fsb}
	f(s,0)=1-\alpha(s)
\end{equation}
and 
\begin{equation}
	f(s,0) =1/2-\alpha(s),\label{fsb0}
\end{equation}
where the function $\alpha(s)$ is positively defined and vanishing at $s\to\infty$. As it has been discussed above, the model estimate for this function presuppose the dependence proportional to $ s^{-\lambda}$.

Saturation of unitarity  leads to the following  $s$-dependencies of the elastic and inelastic overlap functions at $b=0$:
\begin{equation}\label{hesb0}
	h_{el}(s,0)=1-2\alpha(s)+\alpha(s)^2,
\end{equation}
i.e. vanishing of $h_{inel}(s,0)$ at $s\to\infty$ :
\begin{equation}\label{hisb0}
	h_{inel}(s,0)=\alpha(s)-\alpha(s)^2=\alpha(s)(1-\alpha(s))
\end{equation}
Saturation of the black disk limit implies that
\begin{equation}\label{hesb0r}
	h_{el}(s,0)=1/4-\alpha(s)+\alpha(s)^2
\end{equation}
and nonvanishing inelastic overlap function
\begin{equation}\label{hisb0r}
	h_{inel}(s,0)=1/4-\alpha(s)^2.
\end{equation}
Eqs. (\ref{fsb}-\ref{hisb0r}) allow one to trace approaching asymptotic limits for the black disk and unitarity saturations.
The  inelastic overlap functions values at $b=0$  at the three LHC energies are  illustrated at Fig. 1.
\begin{figure}[hbt]
	\begin{center}
	\resizebox{10cm}{!}{\includegraphics*{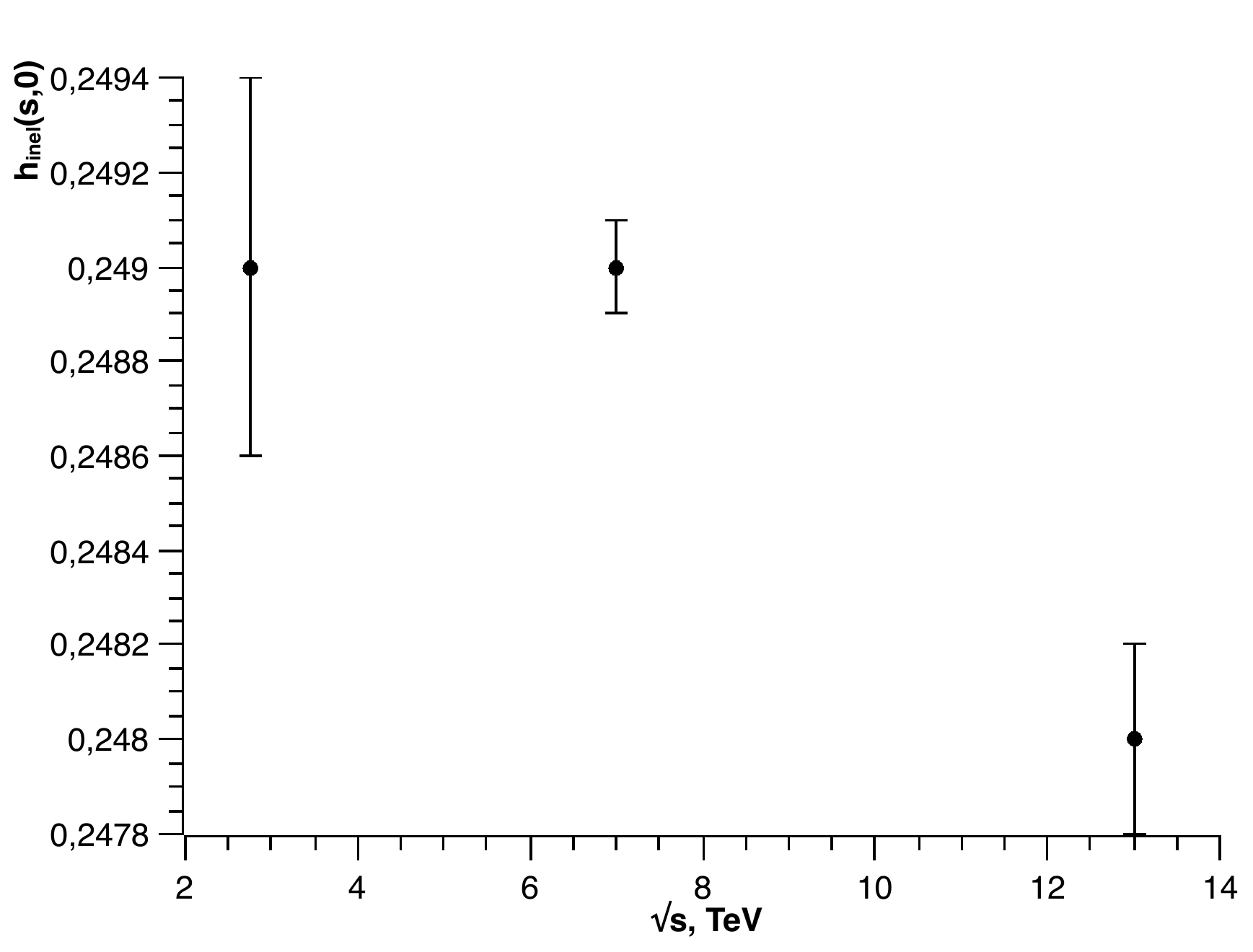}}
	\end{center}
	\hspace{3cm}
	\vspace{-0cm}
	\caption[ch2]{\small Values of the inelastic overlap function  at $b=0$  for the three LHC enegries ($\sqrt{s}=2.76,\, 7,\, 13$ TeV)  The ``experimental'' points are extracted from reference \cite{l2}.}
\end{figure}

Saturation of the above limits leads to vanishing of the real to imaginary parts of the scattering amplitude ratio $\rho(s)$ 
 at $t=0$ when $s\to\infty$ \cite{me,fin}. The unitarity limit means that $h_{inel}(s,0)\to 0 $  at $s\to \infty $. 
 Generic energy depencies of the overlap functions are represented schematically at Fig. 2. 
 \begin{figure}[hbt]
 	\begin{center}
 		\resizebox{15cm}{!}{\includegraphics*{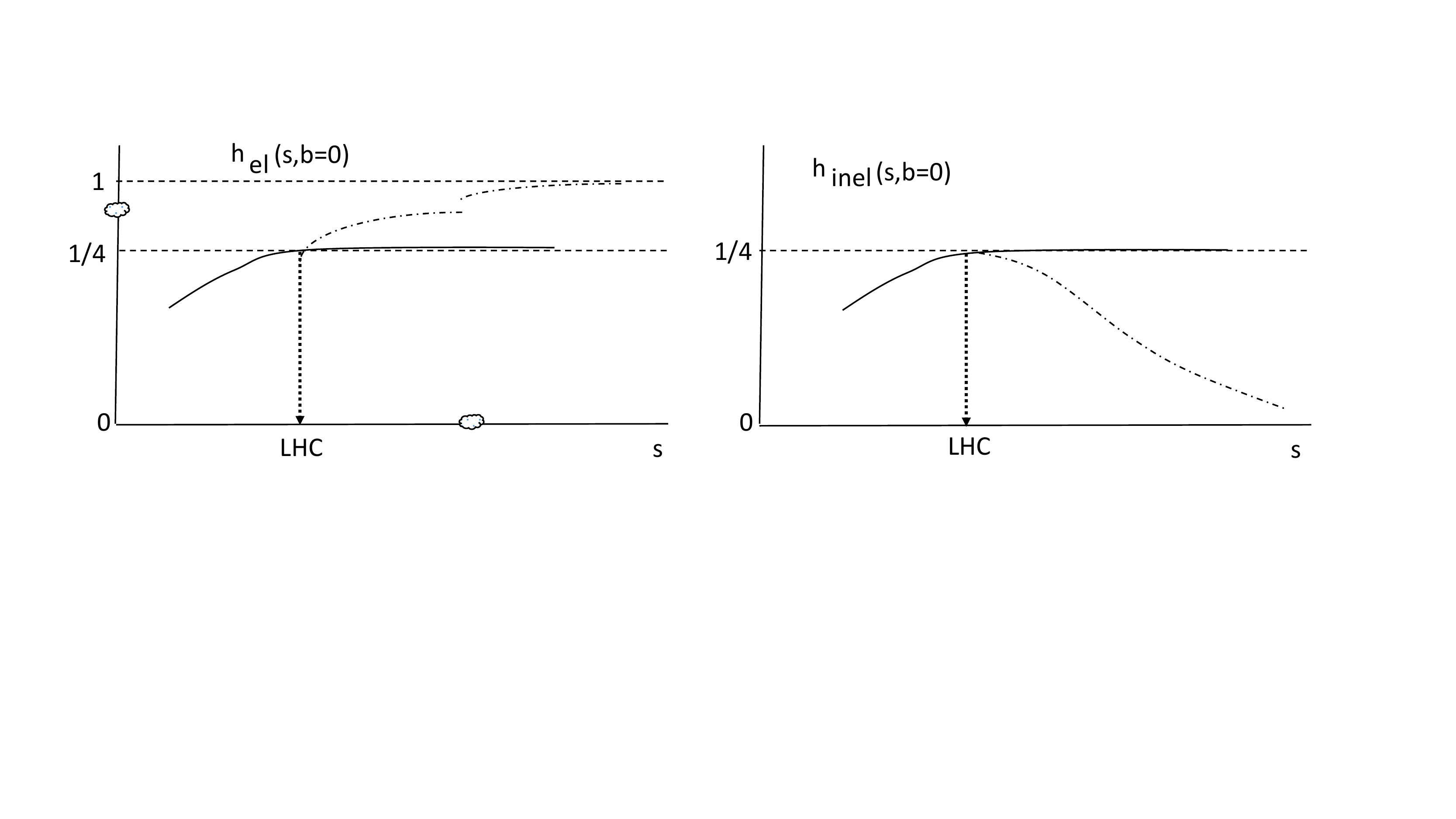}}
 	\end{center}
 	\hspace{3cm}
 	\vspace{-4cm}
 	\caption[ch2]{\small Overall energy dependencies of the overlap functions at $b=0$.}
 \end{figure}
 
  The region of small impact parameter values $b\sim 0$ corresponds to the deep--elastic scattering.
 In this region the ratio of the differential cross--sections   should be equal to 4 asymptotically \cite{del}. Thus, the reflective mode enhances large $-t$ tail of $d\sigma/dt$. Interpretation of this enchancement is associated with the hard  core presence in  hadrons and their interaction under collision  \cite{physl21}. The estimated cores interaction radius constitutes 46 \% of protons interaction  radius and corresponds to the estimates given in \cite{fuku}. 

At the LHC, there is no significant difference between the ``Larger'' components of the BEL and REL regimes, the expectation values  $\langle b^2 \rangle _{el,inel}(s)$ have similar dependence at $s\to\infty$ \cite{epl}.

\section*{Conclusion}
 The experiments at the LHC  have shown a collapse of the BEL mode  since the maximal absorption has been reached or even exeded in the central collisions.
 A speed up of the ratio  $\sigma_{el}(s)/\sigma_{tot}(s)$ increase could be expected then due to transition to the REL regime and redistibution between the elastic and inelastic probabilities due to unitarity, when the maximal probability of inelastic interactions is shifted to the peripheral region  of $b_{max}\sim \mu^{-1} \ln s$. It corresponds to an emergent  black ring with growing radius of the inelastic probability distribution  over the impact parameter. And it is correlated with a presumably observed onset of decrease of  the inelastic overlap function 
 $
h_{inel}(s,0)
 $
 and appearing central reflection at the LHC energy range. Symmetry considerations of  $
 h_{inel}(s,0)
 $
  are in favor of coexistence of absorptive and reflective scattering modes.
  
  Studies of the overlap functions at $b=0$ at the different values of energy in the range of LHC is the main point of this note.
 Such studes of the overall energy dependence of the overlap functions become  possible now due to availability of the data on  elastic scattering in a wide energy range of the LHC.
 Decrease of the inelastic overlap function at $b=0$  is a consequence of  approach to unitarity saturation when $h_{inel}(s,0)\to 0$
 and $\sigma_{inel}(s)/\sigma_{tot}(s)\to 0$ at $s\to \infty$\footnote{Asymptotically $\sigma_{inel}(s)\sim\ln s$, while $\sigma_{tot}(s)\sim\ln^2 s$ in the REL mode,  cf. e.g. \cite{edge}.}.
  Experimentally observed energy dependence of $h_{inel}(s,0)$  can   be  used as a tool for discrimination of the reflective scattering mode  from the absorptive one at the highest available energies and beyond\footnote{We do certainly not imply oscillating energy dependencies for extrapolations to higher energies but expect monotonous approach to the unitarity limit.}. 

 The  updated version \cite{kfkn} of the earlier model analysis \cite{kfk} is in line with the expected transition between BEL and REL modes  and implies that 
 \begin{equation}
 E(s)>0, \,I(s)<0
 \end{equation}
  at $\sqrt{s}=13$ TeV in correspondence with earlier phenomenological considerations \cite{l2,mart}. 
  Thus, the LHC measurements have revealed transition \cite{l2} to the black ring scattering picture \cite{93} inspired  by the Tevatron CDF data \cite{giro}.

 Magnitude of  $X(s)$ depends on intensity of interactions and does not depend on the interaction radius.  The model confirmation of this  statement can be found in \cite{strum}.  Indeed, the same conclusion is valid for the both functions $X(s)$  and $Y(s)\equiv \sigma_{tot}/16\pi B(s)$ \cite{intens}, where $B(s)$ is the forward slope parameter. Saturation of the unitarity limit is correlated with  approach to the Auberson--Kinoshita--Martin (AKM) scaling \cite{akm}  domain where the both ratios $X(s)$  and $Y(s)$ become
 energy--independent and tend to unity at asymptotics.
  
  A commonly accepted statement is that the observed increase of $X(s)$   implies that the interactions at the LHC energies are far from the asymptotics  and  the use of asymptotic theorems  is therefore precocious. The functions $I(s)$ and $E(s)$ can help under the studies of interactions in the preasymptotic energy region and of the respective scattering mode. Availability of  several energy values  for experimentation at the LHC  allows one to get hints on the energy dependence of the functions $I(s)$ and $E(s)$. 
  
  And to extract these functions from data one does not need to use Hankel transform with rapidly oscillating Bessel functions. Since the $I(s)$ and $E(s)$ are defined as  integrals over $\sqrt{-t}$, it minimizes phase--related ambiguity  of the scattering amplitude extracted from the data on $d\sigma/dt$ at large transferred momenta values beyond the dip position. It also overcomes some
  uncertainties in a model description of the large--$t$ region.

  It should be emphasized that transition between BEL and REL modes implies crossing of the black disc limit by the scattering amplitude.
 The energy evolution of the elastic $S$--matrix scattering element  $S(s,0)$, the elastic  scattering amplitude $f(s,0)$  and the overlap functions $h_{el}(s,0)$ and $h_{inel}(s,0)$ starting from the energy $s_i$ to the energy $s_f$
 across the black disc limit energy value 
 \[
 	s_i\to s_{m}\to s_f
 \]
 can be summarized  below  by the several interrelated relations:
 \begin{equation}\label{sii}
 	S(s_i,0)>0 \to S(s_f,0)<0,
 \end{equation}
 \[
 	f(s_i,0)<1/2\to f(s_f,0)>1/2,
 \]
 \[
 	h_{el}(s_i,0)<1/4\to h_{el}(s_f,0)>1/4,
 \]
 \begin{equation}\label{si}
 	h_{inel}(s_i,0)<1/4
 	 \to h_{inel}(s_f,0)<1/4.
 \end{equation}
  A nonmonotonic (loop variation) energy dependence of the inelastic overlap function $h_{inel}(s,0)$   results directly from transition between BEL and REL scattering modes when the function $S(s,0)$ changes its sign from  a positive to  a negative one,  Eq. (\ref{sii}).	Again, such variation implies coexistence ot the two scattering modes.

  Transition from the BEL to the REL regimes of hadron scattering can be considered as a result of a phase transition between a color insulating state of hadron matter and a color conducting one of QCD matter \cite{jpg}. It seems promising therefore to study an energy evolution of the elastic and inelastic overlap functions at the LHC in the head-on hadron collisions where the maximal overlap of 
  hadronic matter takes place in view of quark--hadron duality. Extrapolation of the overlap functions into the energy region beyond of the LHC energies helps in obtaining hints on the nature of the asymptotic scattering dynamics and  properties of  media formed in the hadron interaction region.

\section*{Acknowledgement}
We are grateful to L\'aszl\'o Jenkovszky for  the interesting discussions on the Odderon properties and general remarks. We would like to thank the Rewiewers for the consideration of our manuscript and many useful recommendations.

\end{document}